\documentclass[prl,twocolumn]{revtex4}%
\usepackage{amsmath}
\usepackage{amsfonts}
\usepackage{amssymb}
\usepackage{graphicx}%
\usepackage{setspace}

\begin{document}



\title{Loss compensation in Metamaterials through embedding of active transistor based negative differential resistance circuits}

\author{Wangren Xu$^1$, Willie J. Padilla$^2$, and Sameer Sonkusale$^1$$^\star$}
\affiliation{$^1$NanoLab, Department of Electrical and Computer Engineering, Tufts University, 161 College Ave, Medford, MA 02155, USA}
\affiliation{$^2$Department of Physics, Boston College, 140 Commonwealth Ave., Chestnut Hill, MA 02467, USA}

\email{sameer@eecs.tufts.edu}

\begin{abstract}

This paper presents an all-electronic approach for loss compensation in metamaterials. This is achieved by embedding active-transistors based negative differential resistance (NDR) circuits in each unit cell of the metamaterial lattice. NDR circuits provide tunable loss compensation over a broad frequency range limited only by the maximum operating frequency of transistors that is reaching terahertz values in newer semiconductor processes. Design, simulation and experimental results of metamaterials composed of split ring resonators (SRR) with and without loss compensation circuits are presented.

\end{abstract}

\maketitle

Metamaterials derive their exotic electromagnetic (EM) properties from their artificially engineered structure made of metal and dielectric material instead of constituent properties of atoms and molecules~\cite{Veselago68,Pendry99,Pendry00,Zheludev10}. Metamaterials have shown promise in the general area of transformation optics~\cite{Chen10,Kildishev11} with applications such as negative index of refraction~\cite{Shelby01}, perfect lens ~\cite{Zhang11}, superlens~\cite{Zhou11b} , electrically small antenna ~\cite{Zhu11} and perfect absorbers~\cite{Zhou11a,Hedayati11,Singh11}. Metamaterials suffer from metallic and dielectric losses~\cite{Zheludev10} that eventually limit their overall performance. This issue is exacerbated in the 3D metamaterials where there is more interaction between the electromagnetic waves and the bulk metamaterial ~\cite{Dong10}. Loss mitigation approaches are needed for metamaterial devices to be practical for real applications. One approach reduce the losses by geometric tailoring of individual metamaterial unit cell. For example, G$\ddot{u}$ney \textit{et al.} found that they could reduce the ohmic loss by using bulkier or wider wires, which is equivalent to increasing the fill ratio of unit cells at terahertz frequencies~\cite{Guney09}. This accounts for the fact that skin depth is not negligible compared to lattice parameter at terahertz frequencies. It has also been reported that increasing the inductance to capacitance ratio can reduce the losses of metamaterials~\cite{Zhou08}. However, these methods restrict the geometric structures allowed for metamaterial design. Approaches that suggest means to compensate for losses are more desirable.  One such approach for loss compensation is to introduce a gain medium into metamaterials~\cite{Dong10,Stockman11,Xiao10,Fang11,Ramakrishna03,Plum09}. However, this scheme requires optical pumping that consumes power and limits compensation over a narrow frequency band. It was also noted in ~\cite{Soukoulis10} that optical pumping methods for loss compensation may be highly impractical and unscalable. Instead, an electrical approach for loss compensation that works by injecting electrical current from a semiconductor into a chunk of metamaterial may be more suitable ~\cite{Soukoulis10}. One example of an electrical method for loss compensation recently shown for split ring resonators (SRR) is to insert additional smallers SRRs with an amplifier signal chain to sense, amplify and inject amplified signal in to the metamaterial to compensate for loss~\cite{Popa07,Yuan09}. However, additional antenna elements affect the original metamaterial response and adds design complexity especially at shorter wavelengths when lattice dimensions get smaller.

\begin{figure*}[!]
\begin{center}
\includegraphics[width=\textwidth,height=\textheight,keepaspectratio=true]{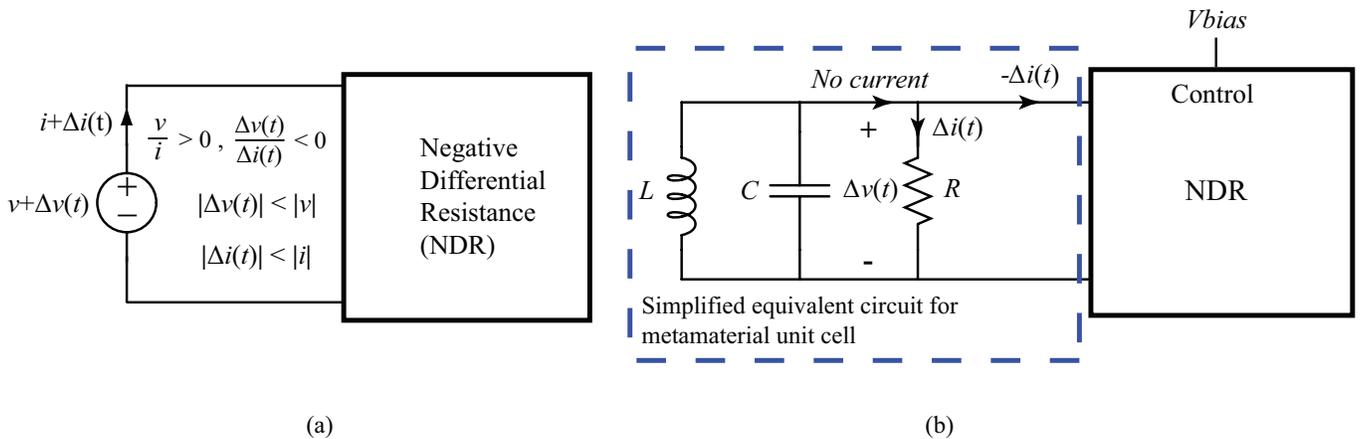}
\caption{(a) The definition of NDR (b) simplified equivalent circuit model: The SRR is modeled by an RLC parallel circuit, in which the $R$ stands for the entire loss of the SRR resonator. Negative differential resistance (NDR) models the active loss compensation circuit seen by the SRR as means for loss compensation. With the loss compensation circuit at the resonant frequency, the variation of  current flowing into metamaterial loss $R$ and the variation of current flowing into NDR have the same amplitude $\Delta i(t)$ but opposite polarity. The NDR can be tuned by bias voltage $Vbias$ externally}
\label{Fig1}%
\end{center}
\end{figure*}

In this paper, we present a novel active-transistor embedding approach for loss compensation. The idea is to incorporate active transistor based integrated circuits directly into each metamaterial unit cell. The embedded circuitry provides a negative differential resistance (NDR) to compensate for the loss in each unit cell. The concept of NDR is explain in Fig. \ref{Fig1} (a).  By definition, NDR circuit exhibits a negative value for impedance  $\frac{\Delta v(t)}{\Delta i(t)}$, while keeping a positive value for large scale impedance $\frac{v}{i}$, where $v$ and $\Delta v(t)$ are the DC and small signal voltage across the network, and $i$ and $\Delta i(t)$ represents the DC and small-signal current flowing into the NDR network, respectively. As shown in Fig. \ref{Fig1} (b), to the first order, a metamaterial unit cell can be viewed as a resonant RLC tank circuit with resistance $R$ modeling the undesirable ohmic and dielectric loss. At resonance frequency of the metamaterial, $\omega = \frac{1}{\sqrt{LC}}$, the strength of the resonance is affected by its loss $R$. This also affects the $(\epsilon, \mu)$ parameters of effective medium. One can view the incident EM radiation as inducing a differential voltage $\Delta v(t)$ across the RLC network. The current through the loss resistance $R$ is given by $\Delta i(t) =\frac{\Delta v(t)}{R}$. Connecting a NDR circuitry in parallel with the metamaterial RLC circuit will inject a current $\Delta i(t) = -|\frac{\Delta v(t)}{R}|$ that essentially compensates for current through the lossy resistor $R$. Thus, NDR provides an effective negative differential resistance of value $-R$ at the resonant frequency. While this concept of negative differential resistance (NDR) has been applied here for metamaterials for the first time, it has been widely used in circuits community. For example, oscillations in Voltage Controlled Oscillators (VCO) are prevented from decaying out using negative differential resistance circuits~\cite{Craninckx97,Razavi11}. The NDR circuits can be implemented using very few active and passive components that can all be potentially integrated into a monolithic chip with size negligible compared to the metamaterial unit cell and and with low power dissipation. Aggressive scaling of the semiconductor technology to nanometer dimensions where transistors are reaching maximum oscillation frequencies $f_{max}$ and transit-time frequencies $f_{T}$ of terahertz values ~\cite{Albrecht11,Albrecht10} allows for such a prospect. A terahertz modulator based on embedding of GaAs HEMT in metamaterial was recently demonstrated~\cite{Shrekenhamer11}. This approach can be combined with current approaches based on gain medium that have been applied at infrared and visible frequencies to achieve loss compensation in metamaterials over a broad frequency range.

In this paper, we propose an electronic approach for loss compensation and demonstrate its utility for a class of planar metamaterials based on single-split Split Ring Resonator (SRR). However the approach is valid for any metamaterial construct that are based on patterned metallic inclusions in the dielectric medium. The SRR structure couples to both electric and magnetic field~\cite{Schurig06} and have been a very popular choice with practical demonstrations of negative permeability ~\cite{Shelby01}, electrically small antennas ~\cite{Ziolkowski02} and even superlens ~\cite{Fang03}. Their performance is limited by the ohmic and dielectric losses. The design, simulation and implementation of the SRR based metamaterial with and without loss compensation is discussed below.

\begin{figure}[!]
\begin{center}
\includegraphics[width=2.75in,keepaspectratio=true]{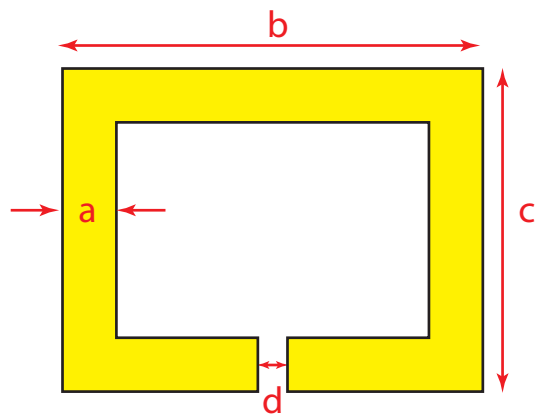}
\caption{(Color online) Dimension of SRR in the test. All units are in mm. a=2, b=16, c=12, d=1.5}
\label{Fig2}%
\end{center}
\end{figure}

The dimension of the original SRR is shown in Fig. \ref{Fig2}. The metallic split ring is made of 1.4 mil thick copper layer. The dielectric substrate is a 62 mil FR4 substrate. The real part of its permittivity is 4.8 and the loss tangent is 0.027 at 2GHz. The equivalent circuit model of SRR can be characterized at the gap of the SRR by measuring the reflection coefficient. The SRR was simulated using a commercial finite difference time domain (FDTD) solver Microwave Studio by CST ~\cite{CST11}.  For experimental verification, SRR was excited by the loop antenna in a configuration similar to ~\cite{Wang08}, which is discussed later.

\begin{figure}[!]
\begin{center}
\includegraphics[ width=2.75in,keepaspectratio=true]{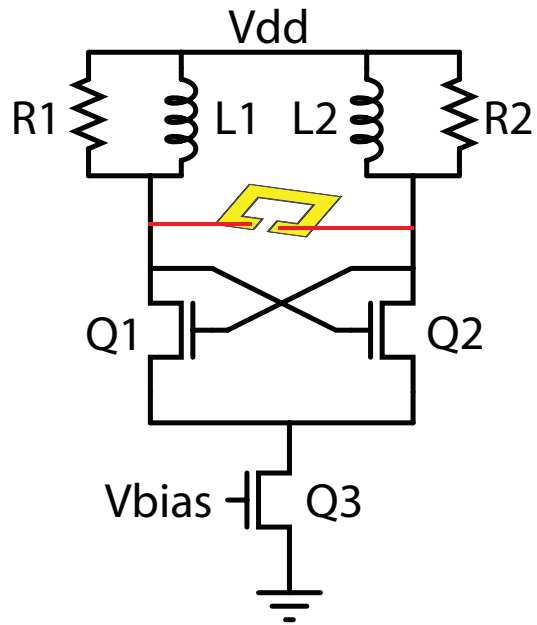}
\caption{(Color online) Implementation of active negative differential resistance circuit and its connection to SRR.}
\label{Fig3}%
\end{center}
\end{figure}

\begin{figure}[!]
\begin{center}
\includegraphics[width=2.75in,keepaspectratio=true]{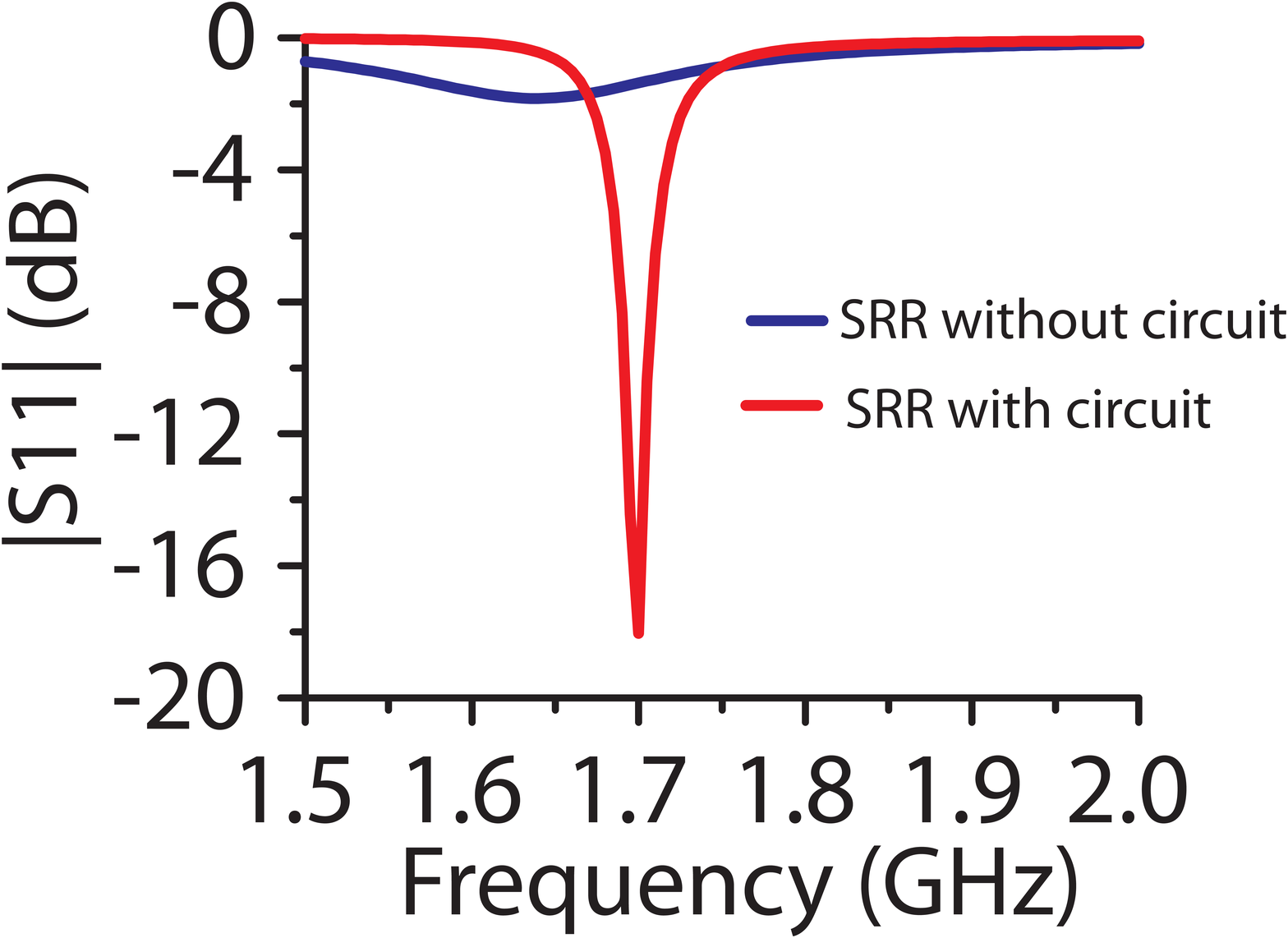}
\caption{(Color online) simulated $\mid{S11}\mid$ in dB of SRR with and without active loss compensation circuit}
\label{Fig4}%
\end{center}
\end{figure}

The implementation of the negative differential resistance (NDR) circuit and its connection to the SRR is shown in Fig. \ref{Fig3}. It consists of just 3 transistors, 2 resistors and 2 inductors. In principle, this circuit is a cross-coupled transistor pair consisting of Q1 and Q2,  L1 and L2 are identical high Q inductors and R1 and R2 are identical resistors to bias the transistors Q1 and Q2. Q3 is a current sink whose current can be tuned by its gate voltage $Vbias$. In our implementation, Q1, Q2  and Q3 were pseudomorphic high electron mobility transistors (pHEMT) and could also have been metal oxide semiconductor field effect transistors (MOSFET) or bipolar junction transistors (BJT). The active negative differential resistance circuit can be regarded as a one-port network looking into the drains of transistor Q1 and Q2 and connected across the gap of the SRR. When appropriately biased, the circuitry imparts negative differential resistance to small signal variations across the gap, equal and opposite in polarity to the loss $R$. The negative differential resistance for this cross-coupled NDR circuitry can be approximated by $-\frac{2}{g_{m}}$, where $g_{m}$ is the small-signal transconductance of Q1 or Q2 and is related to their aspect ratio and bias current. Transconductance $g_{m}$ for a transistor is defined as $\frac{\Delta i}{\Delta v_{control}}$, where $\Delta i$ is the small signal current between the two terminals of the transistor in response to the small signal variation $\Delta v_{control}$ at the gate (controll/third terminal) of the transistor ~\cite{transistor}. Therefore, we can tune the negative differential resistance by changing the aspect ratio of transistors during the design phase or by tuning the bias voltage of current sink Q3. The active negative differential resistance circuit also provides a reactive component for the SRR, whose value can be tuned by changing L1 and L2 in Fig. \ref{Fig3}. For demonstration, we simulated the circuit in Fig. \ref{Fig3} in Advanced Design System (ADS) by Agilent ~\cite{ADS11}. By running an EM-circuit co-simulation in CST and ADS, we obtain the simulation result of reflection $|S11|$ in dB of SRR as shown in Fig. \ref{Fig4}. Before the active circuit is connected, the resonant frequency of SRR is around 1.68 GHz and the magnitude of its reflection at this frequency is around -2 dB. After the active loss compensation circuit is connected to the SRR, the resonant frequency is shifted slightly to 1.7 GHz and the magnitude of S11 is increased to around -18 dB. The slight frequency shift is due to process variations and presence of parasitic capacitances for transistors Q1 and Q2. This change can be incorporated by pre-embedding the parasitics during the design phase (we added a 0.5 pF capacitor to account for this shift in order to center the frequency back to 1.68 GHz).


There are many design considerations in embedding of active transistor circuit elements into SRR, the most important being the stability of the metamaterial and the circuit providing negative differential resistance. The active negative differential resistance circuit (at port 3) connected across the  gap of the SRR ( at port 2) forms a closed loop system with a loop gain $$A_{closed\ loop}=|S22_{SRR}|*|S33_{active\ circuit}|$$
According to \emph{Barkhausen stability criterion}~\cite{Bongiorno63}, the system is stable as long as the loop gain $A_{closed\ loop}$ is below unity for a total phase shift of $2\pi$ across the loop. This can be guaranteed by monitoring the onset of self-oscillations (the positive feedback) without external excitation, and then adjusting the current through transistors Q1 and Q2 via $Vbias$ control of Q3 in Fig. \ref{Fig3} to eliminate them. Ideally, the reactive component of active negative differential resistance circuit cancels that of the SRR, which leaves an almost negligible overall loss at the resonant frequency.

Next we provide measurement results to verify this concept of embedding active transistor circuits into metamaterial for loss compensation. We have implemented a Split Ring Resonator (SRR) embedded with loss compensation circuit on a PCB with the parameters discussed earlier. The dimension of the SRR is shown in Fig. \ref{Fig2}. The resonance frequency of the SRR is around 1.7 GHz. The ratio between the wavelength and the SRR size is as large as 11.03. The circuit topology of the loss compensation circuit is the same as that shown in Fig. \ref{Fig3}. Q1, Q2 and Q3 are implemented using Avago ATF-54143 low noise enhancement mode pseudomorphic HEMT(pHEMT) transistors in a surface mount plastic package. L1 and L2 are high-Q 1nH surface mount inductors. R1 and R2 are surface mount 100$\Omega$ resistor. We added more loss by introducing 2 surface mount 1$\Omega$ resistors in series at the two ends of the SRR.

\begin{figure}[!]
\begin{center}
\includegraphics[width=2.75in,keepaspectratio=true]{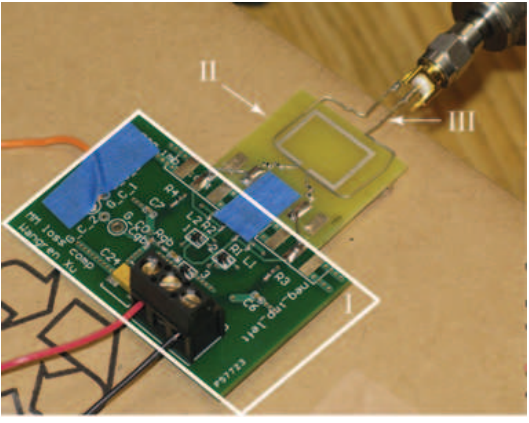}
\caption{(Color online) Measurement setup to measure the reflection. Area I shows the active negative differential circuit implemented on the printed circuit board. Area II shows the SRR. Area III shows the loop antenna. The circuit in area I could be integrated into single chip in future, with dimensions much smaller than the size of metamaterials.}
\label{Fig14}%
\end{center}
\end{figure}

Excitation was provided using a loop antenna as proposed in~\cite{Wang08}. We made the loop antenna with tinned copper wire, slightly larger than the SRR and placed at a distance of 2 mm over the SRR. Measurements were taken with vector network analyzer Agilent HP8510C.  The loop antenna has a self-resonant frequency at 3.9 GHz, much higher than the expected metamaterial resonance. The SRR with loss compensation circuitry is mounted underneath the loop antenna and an expected resonance at around 1.7GHz in $|S11|$ is measured. The bias voltage of the current sink Q3 of the active negative differential resistance circuit is set to several increments between 0V (no loss compensation) and 0.351V ( maximum loss compensation). The current in the active circuit is measured to be just 3.2 mA. The measurement setup is shown in Fig. \ref{Fig14}. As we sweep the bias voltage in steps between 0V to 0.351V, the resulting $|S11|$ in dB shown in Fig. \ref{Fig15} indicates reduction in overall loss by gradual increase in the strength of the resonance. The measurement shows that  resonance strength has been improved from -8dB to -32dB, which also agrees well with simulation in Fig. \ref{Fig4}. The resonant frequency was shifted slightly because of the parasitics of the active and passive components.

\begin{figure}[!]
\begin{center}
\includegraphics[width=2.75in,keepaspectratio=true]{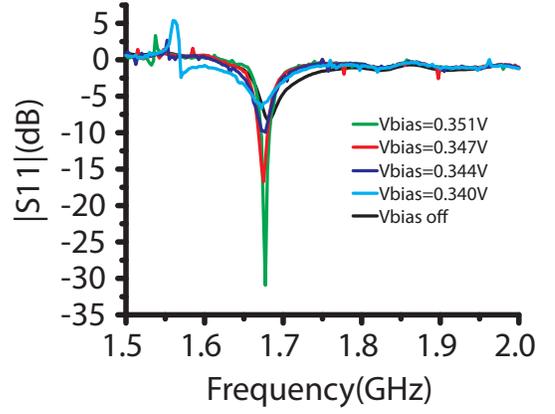}
\caption{(Color online) Measured S11 of SRR when we sweep the bias voltage of the current sink in the active negative differential circuit. When the bias voltage is 0.351V, the current flowing in the active circuit is 3.2mA. When the bias voltage is decreased to 0.347V, the current goes down to 3mA. When the bias voltage is 0.344 V, the current becomes 2.9 mA. When the bias voltage is 0.340V, current is 2.5 mA}
\label{Fig15}%
\end{center}
\end{figure}

We also retrieved the effective permeability of the SRR simulated with plain wave excitation in CST using the electromagnetic parameter retrieval method proposed in~\cite{Smith05}. The equations~\cite{Smith05} used to retrieve the effective permeability are:
\begin{eqnarray}
n=\frac{1}{kd}\arccos[\frac{1}{2S_{21}}(1-S^{2}_{11}+S^{2}_{21})]\\
z=\sqrt{\frac{(1+S_{11})^{2}-S_{21}^{2}}{(1-S_{11})^{2}-S_{21}^{2}}}\\
\mu=z\cdot n
\end{eqnarray}
where $n$ is the refractive index, $z$ is the wave impedance of the metamaterial slab, $d$ is the thickness of the slab (18 mm), $k$ is the wave number of the incident wave, $S11$ is the reflection coefficient and $S21$ is related to the transmission coefficient $T$ by $S21=Te^{ikd}$ ~\cite{Chen04}. The extracted permeability of original SRR and SRR embedded with loss compensation circuit is shown in Fig. \ref{Fig15}. With the embedded loss compensation circuit, the resonance of both the real and imaginary part of the permeability become stronger and sharper around the resonant frequency indicating loss compensation in metamaterials.

\begin{figure}[!]
\begin{center}
\includegraphics[width=2.75in,keepaspectratio=true]{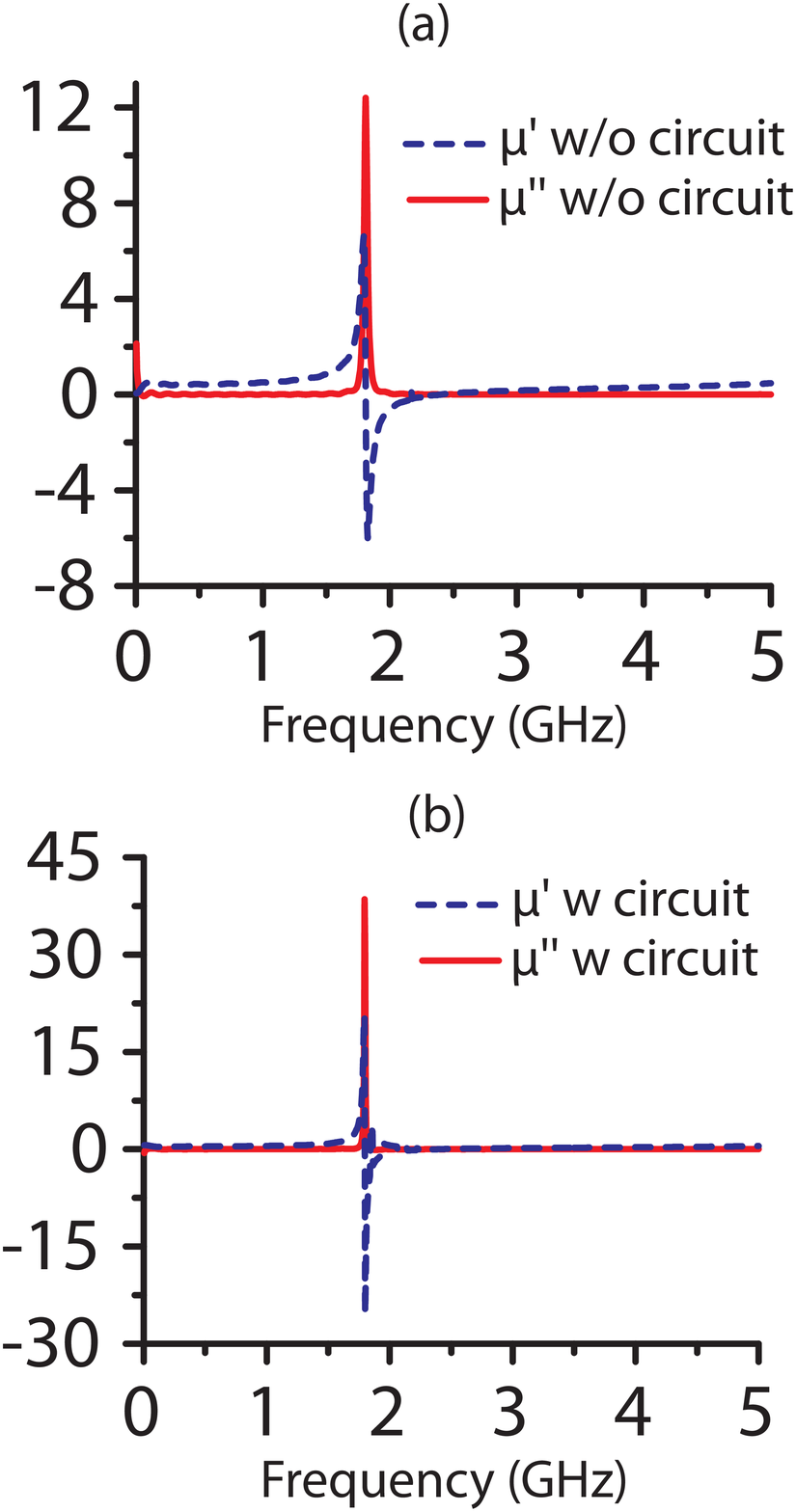}
\caption{(Color online) Retrieved complex permeability $\mu=\mu^{\prime}+i\cdot\mu^{\prime\prime}$ of SRR excited by plane wave (a)original SRR (b)SRR embedded with loss compensation circuit }
\label{Fig15}%
\end{center}
\end{figure}


In conclusion, this paper provides a method of loss compensation in metamaterials through embedding of active transistor based negative differential resistance circuit directly into each metamaterial unit cell. Because of the negligible size of the embedded circuit, it does not interfere the EM response of the metamaterial. It was also shown that the negative differential resistance value can be tuned off-chip and this will allow one to guarantee stability of the circuitry and also accommodate for any process and environmental variations. The resonance frequency of metamaterial can also be tuned through additional on chip varactors in the active negative differential resistance circuit. As the semiconductor technology scales to nanometer dimensions with maximum oscillation frequency exceeding terahertz, proposed electrical method for loss compensation integrated into a single chip of dimensions much smaller than the metamaterial lattice parameter. Also it can be applied to metamaterials working at Radio Frequency (RF), microwave, millimeter-wave and terahertz frequencies. Since continued transistor scaling also comes with reduced power dissipation, the overall power consumption for loss compensation circuit will be amenable for practical implementations. The proposed loss compensation can be used to make ideal absorbers, cloaks, perfect lens, modulators, phase shifters, electrically small antennas, etc. with applications in radar imaging, cellular mobile communications, satellite communications, antennas and security screening.

\end{document}